\documentclass[journal]{IEEEtran}
\usepackage[pdftex]{graphicx}
\usepackage{amsmath, amssymb}

\usepackage{tabularx,booktabs}
\usepackage{cite}
\usepackage[ruled,vlined]{algorithm2e}

\usepackage{algorithmic}

\DeclareMathOperator{\E}{\mathbb{E}}

\makeatletter
\newcommand\fs@norules{\def\@fs@cfont{\bfseries}\let\@fs@capt\floatc@ruled
  \def\@fs@pre{}%
  \def\@fs@post{}%
  \def\@fs@mid{\kern3pt}%
  \let\@fs@iftopcapt\iftrue}
\makeatother

\ifCLASSINFOpdf
  
\usepackage{xcolor}
\AtBeginDocument{\colorlet{defaultcolor}{.}}

\begin{document}

\title{A Deep Reinforcement Learning-Based Caching Strategy for IoT Networks with Transient Data }

\author{Hongda~Wu,~\IEEEmembership{Student Member, IEEE},  Ali~Nasehzadeh,~\IEEEmembership{}
        Ping~Wang,~\IEEEmembership{Fellow,~IEEE}
\thanks{Hongda Wu, A. Nasehzadeh, and Ping Wang are with the Department
of Electrical Engineering and Computer Science, York University, Toronto,
ON, M3J 1P3 Canada e-mail: hwu1226@eecs.yorku.ca, arian92@eecs.yorku.ca, pingw@yorku.ca.}
}

\maketitle

\begin{abstract}
The Internet of Things (IoT) has been continuously rising in the past few years, and its potentials are now more apparent. However, transient data generation and limited energy resources are the major bottlenecks of these networks. Besides, minimum delay and other conventional quality of service measurements are still valid requirements to meet. An efficient caching policy can help meet the standard quality of service requirements while bypassing IoT networks’ specific limitations. Adopting deep reinforcement learning (DRL) algorithms enables us to develop an effective caching scheme without the need for any prior knowledge or contextual information. In this work, we propose a DRL-based caching scheme that improves the cache hit rate and reduces energy consumption of the IoT networks, in the meanwhile, taking data freshness and limited lifetime of IoT data into account. To better capture the regional-different popularity distribution, we propose a hierarchical architecture to deploy edge caching nodes in IoT networks. The results of comprehensive experiments show that our proposed method outperforms the well-known conventional caching policies and an existing DRL-based solution in terms of cache hit rate and energy consumption of the IoT networks by considerable margins.

\end{abstract}


%
\IEEEpeerreviewmaketitle

\section{Introduction}
%
%
%
%
\IEEEPARstart{T}{he} world is flooding with data and data transmissions; simultaneously, the demand for more reliable and faster connections is souring. This concurrency results in either terribly congested networks or the need for more resources and more intelligent networking schemes. With the emergence of the Internet of Things (IoT), there is an exponential growth in the number of connected devices in the world, which again emphasizes the importance of better networking schemes \cite{myConf} \cite{iotStuff}. Caching is a networking technique that enables networking nodes to store frequently requested files, mitigating network traffic and improving the response time. In cases where the downlink is more heavily used, such as in content delivery networks (CDNs), extremely large files need to be transmitted to users. Caching can prevent multiple transmissions from the source by responding to users from the edge nodes where the files have been downloaded earlier. \textit{Edge caching} is a new technology that enables edge nodes, e.g., base stations or user's devices, to be a part of the caching schemes and store files. Being close to the end-users means that the request does not necessarily go all the way up to the source to fetch the response because one of the edge nodes might already have the desired file in its caching memory and is ready to fulfill the request, which in turn will significantly shorten the response time and simultaneously ease the load on the backhaul link \cite{6225433}.

Numerous caching methods have been proposed previously; some well-known methods include Least Frequently Used (LFU) and Least Recently Used (LRU) methods \cite{lfu}. The basic idea for both caching algorithms is to rank the files in the nodes’ caching memories. The lowest-ranked file will be replaced with the new one if the node has to store a new file that was not stored before. The difference between these two methods lies in how they approach the ranking phase. Though the effectiveness is shown on conventional caching problem, both policies have several drawbacks \cite{ccVelipasalar, 8308415}, especially for handling IoT caching problem where the data lifetime is limited and device energy is constrained. In addition, a massive of IoT devices in the network makes this decision-making problem even more difficult \cite{7883826}.

With the promising advancements in artificial intelligence and machine learning in recent years, Reinforcement Learning (RL) \cite{textbookRL}, a machine learning paradigm, has recently gained significant attention. The goal of RL algorithms is to achieve an optimal policy that leads to a maximized benefit (a.k.a. cumulative reward) that can be utilized for control/optimization purposes. After an RL agent iterates the observation-interaction cycle numerous times, the agent will develop a sequence of mapping from states to actions, which is viewed as optimal policy. However, due to the limited representative capability of Q-table in conventional RL, one can apply RL methods only to the problems with low-dimensional searching space or the cases where the main features could have been easily handcrafted. Build on deep learning techniques, Deep Reinforcement Learning (DRL) unfolds countless possibilities to address problems with huge searching space \cite{Qlearning}. Without providing prior knowledge of networks (e.g., popularity distribution of files) and explicitly defining essential features of learning problems, instead, DRL entrenches the learning process by numerous rounds of interactions with the environment and therefore is viewed as an excellent approach to solve caching problems \cite{zhu2018deep}.

Many existing works have applied DRL to caching problems for different scenarios, including Content Delivery Network (CDNs) \cite{ccVelipasalar, sadeghi}, mobile networks  \cite{vanet3d,pre7Iintegratedbigdata}, and IoT networks\cite{9310745, pre8Jointfog, GreenDRL, QoEDRL, ma2020age, 8713465, 9144224, federatedDRL, mainref}. Nevertheless, most of them
did not take the limited data lifetime and constrained device energy into account. To make IoT caching problem more practical, we focus on the freshness feature of IoT data and take energy consumption into consideration. In addition, we propose a hierarchical architecture for caching rather than using the conventional single-layer architecture to better capture the heterogeneous popularity distribution in different regions. Our main contributions in this paper are as follows.
\begin{itemize}
   \item We formulate the IoT caching problem as a Markov decision process (MDP) while considering real restrictions in this scenario, i.e., limited lifetime of caching data and constrained energy of IoT devices. The newly defined reward function incorporates the data freshness, so as to better imitate the real IoT networks when designing the DRL solution. We adopt the proximal policy optimization solver to get the optimal policy.
   \item We propose a more practical architecture for deploying the caching node in IoT networks. The proposed two-layer hierarchical architecture (i.e., parent-leaf node architecture) takes regional-different popularity distribution into consideration, aside from being more realistic, it also leads to a better system performance.
    \item We conduct extensive experiments, gather comprehensive results, and compare the proposed caching strategy with different caching methods and architectures. The experimental results indicate that the proposed caching strategy can improve the cache hit rate and reduce energy consumption compared to the conventional caching schemes and existing DRL solution \cite{mainref}. Compared with existing benchmarks, the proposed algorithm also demonstrates superior performance in handling the limited lifetime of data.
\end{itemize}

\section{Related Works}
Existing works have leveraged DRL algorithms to find a good caching policy for CDNs with huge files \cite{ccVelipasalar, sadeghi} and mobile networks \cite{vanet3d,pre7Iintegratedbigdata}. In particular, the authors of \cite{sadeghi} considered a hierarchical system model consisting of parent and leaf nodes. They developed a cooperative caching scheme where each node’s caching decision is affected by the decisions of others. The proposed Hyper Deep Q-Network (HDQN) requires stacks of deep Q-networks (DQN) to produce a caching decision. The output of DQNs forms a cost vector, which leads to the DRL agent returning a caching decision. In \cite{ccVelipasalar}, the authors considered the scenario that multiple edge nodes are cooperative and competitive. The proposed Wolpertinger is based on actor-critic architecture with a shrunken action space by the K-nearest neighbor algorithm. The smaller action space allows the algorithm to have a faster run time without significant performance degradation. The authors of \cite{vanet3d} proposed a DRL-based caching strategy for vehicular networks leveraging the multi-view videos from street cameras to proactively cache the required contents and deliver a higher video quality. The term “proactively” refers to the caching policy in which a networking node can ask for a file and store the file in its caching memory, just because it is anticipating to receive the file request in the near future instead of actually receiving the request. The authors of \cite{pre7Iintegratedbigdata} proposed a joint computation offloading, caching, and resource allocation optimization by taking advantage of mobile social networks’ data. The main feature of their work is the definition of a trust index based on the interaction history among network users. This index indicates which user can share its resources, e.g., memory, to help with caching. 

While the majority of related works concentrate on the caching problem in the context of mobile cloud networks or CDNs, the exponential growth of IoT devices has brought the corresponding caching problem into the spotlight \cite{9310745, pre8Jointfog, GreenDRL, QoEDRL, ma2020age, 8713465, 9144224, federatedDRL, MultiAgentCc, mainref}. Some existing works considered one node, e.g., router or base station, to function the caching behavior \cite{9310745, mainref}, while another series of works involve multiple edge nodes to formulate the IoT caching problem \cite{pre8Jointfog, GreenDRL, QoEDRL, ma2020age, federatedDRL, 8713465, 9144224}. In addition, based on the way to solve IoT caching problem, the latter works can be categorized into two streams, i.e., non-cooperative \cite{pre8Jointfog, GreenDRL, QoEDRL, ma2020age} and cooperative \cite{8713465, 9144224, federatedDRL} edge caching, depending on whether there is a cooperation between multiple nodes with caching capability.  

Nath \textit{et.al.} \cite{9310745}
and the authors in \cite{pre8Jointfog} considered a joint optimization for resource allocation, computation offloading and caching for IoT networks using a DRL-based solution. The former considers the scenario of one caching node and the latter scheme involves multiple caching nodes. The authors of \cite{GreenDRL} brought a new point of view to the caching problem; they used a DRL algorithm to assign proper caching capacity and transmission rate to the nodes in a content-centric IoT network. In \cite{QoEDRL} the authors proposed a DRL-based caching mechanism for a content-centric IoT network, in the meanwhile, focusing on the quality of experience.  

Different from single caching node and multiple nodes without cooperation, authors in \cite{8713465, 9144224} considered the multiple edge nodes for caching and assumed that multiple edge nodes can exchange the information among those edge nodes \cite{8713465} or with an additional central server \cite{9144224}. In another recent work, Wang \textit{et.al.} \cite{federatedDRL} deployed a federated deep reinforcement learning method in order to realize cooperative edge caching. Federated learning trains a global model across multiple decentralized edge devices holding local data samples and shares information (via neural network model) in a privacy-preserved way. The proposed method outperforms the conventional methods such as LRU and LFU, and it achieves similar performance to the centralized DRL approach with a low performance degradation.

However, the newly introduced features of IoT networks bring new challenges to the IoT caching problem, one of which is the periodic data generation by IoT devices. Considering the scenario where an IoT sensor might broadcast the room temperature every five minutes, meaning that the data has a limited lifetime in this scenario, which also indicates the ephemeral significance of those data. Thus, data freshness must be taken into account for caching in IoT networks, which is not satisfactorily reflected by the above mentioned works \cite{9310745, pre8Jointfog, GreenDRL, QoEDRL, 8713465, 9144224, federatedDRL}. Furthermore, energy consumption is also problematic in IoT networks. Because IoT devices are generally battery-powered and hence, periodically retrieving data from them causes the device to leave its energy-saving mode, generate the data, transmit the data, and eventually go back to the energy-saving state. It is obvious that iterating through such a cycle consumes significantly more energy than just resting in the energy-saving state \cite{rana2021systematic}.

In this work, we focus on the data transiency in IoT networks and take the limited lifetime and data freshness into consideration. Ma \textit{et.al.} proposed a framework that utilizes information from the queues of user requests for different dynamic contents that need to be served in the near future \cite{ma2020age}, where the indicator adopted is the age of information (a similar concept to data freshness). Differently, we argue that files are valid for a short period of times and their values can be constantly changing in dynamic networks; this proactive caching does not offer enough value for IoT networks. Another similar work \cite{mainref} mentioned the data transiency in IoT networks. To the best of our knowledge, \cite{mainref} is the only work that simultaneously applies DRL and considers the limited lifetime and data freshness to find a solution to IoT caching problem. On the other hand, there is no mention of the energy consumption rate or how it is affected by the caching policy. Moreover, the system model considered in \cite{mainref} comprises only one networking node (a single router). Even though it is an innovative work, the scalability of such a model is of question.
In this work, aside from focusing on the freshness of IoT data, we also propose a more practical architecture for deploying the caching nodes in IoT networks, which is a hierarchical (i.e., parent-leaf) caching system. Assume a file that is not the most popular one in any location, but it has moderate popularity across all of the regions; in this case, none of the leaf nodes see any reason to cache this file, but the parent node who can observe this general popularity will cache this file and improve the cache hit rate significantly.

The rest of the paper is organized as follows. Section \ref{sysmodel} elaborates on the details of the newly-introduced hierarchical system model. In Section \ref{probfor}, we formulate the IoT caching task as an MDP problem and define the state and reward function for DRL training, considering the lifetime of IoT data. Section \ref{ppo} describes the proximal policy optimization solver that solves the MDP problem. Section \ref{result} contains various experimental results that demonstrate the performance of the proposed method in comparison with other existing caching schemes. Finally, Section \ref{conclu} concludes this work.

\section{System Model}
\label{sysmodel}
\subsection{Network Architecture}
\begin{figure}[!t]
\centering
\resizebox{0.45\textwidth}{!}
{\includegraphics{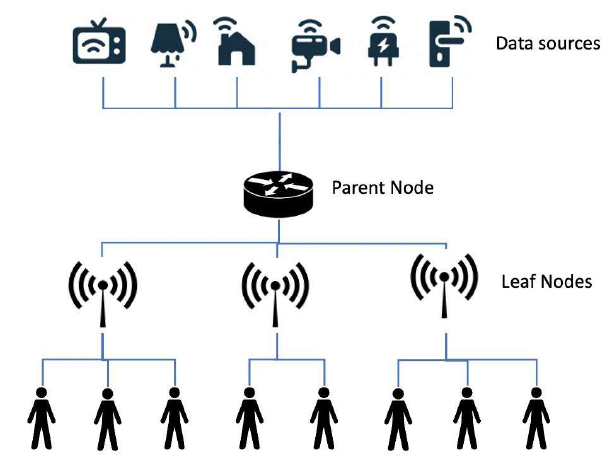}}
\caption{System Model}
\label{Gr: sys_model}
\end{figure}

In real-world IoT network architectures, there may exist multiple layers in a network. In this work, we consider a model consisting of two layers with a parent node at the upper layer and multiple edge/leaf nodes at a lower layer, all capable of caching\footnote{The proposed hierarchical caching architecture can be viewed as a variant of the general one-layer architecture with multiple edge nodes. Since the popularity distribution (see detailed illustration in Section \ref{popdis}) across different regions is different, one advantage behind the proposed architecture is that the parent node can capture the file popularity (with request) in the sense of global view, which may not be reflected by all regional/local edge nodes in one layer. However,
assumptions on parent/edge nodes, i.e., parent node has sufficient storage space to cache all content files (e.g., \cite{MultiAgentCc}) and multiple edge nodes can exchange the information (e.g., \cite{8713465, 9144224}), are not needed. The empirical results in Section \ref{result} show the effectiveness of the proposed architecture when an equal sum of memory space is dispatched.}. Figure \ref{Gr: sys_model} depicts the general architecture based on which we have implemented our work. There are different entities in this IoT network, namely IoT devices, parent node, leaf (edge) nodes, and users. In what follows, we briefly introduce each of them.
\begin{itemize}
   \item IoT Devices: These devices are deployed in the physical world, and they are supposed to sense their environment or control a part of it. Light sensors and smart locks are two examples of such devices. As an inevitable part of their functionality, IoT sensors produce some type of data, which users ask for. One specific attribute to IoT systems is that the data produced by respective devices have a \textit{limited lifetime}, meaning that the data is only valid for a limited time duration after their generation. This is a challenging characteristic to handle for data caching. For instance, a popular file might have an extremely short life span, making it practically non-cacheable.
   
   \item Parent Node: This type of node shapes the inner layers of a network, and it acts as a gateway that connects all the IoT devices at one side and all the edge nodes at the other side. The parent node receives the data requests from the edge nodes, fetches the data from the IoT devices, and then forwards the data to the edge nodes. 
   
   \item Leaf Nodes: These are the networking nodes placed close to the end users. Their coverage is limited to one region with some users, and they are connected to the parent node.  
   
   \item Users: Users could be the applications on some devices such as personal computers, smartphones, or even air conditioning machines. These applications request the data generated by the IoT devices and use them for analysis or control objectives, e.g., a user can monitor the room temperature with a smartphone application using data generated by an IoT sensor. Owing to the short request response time, it is reasonable to assume limited movement for the users and thus a stable connection between users' devices and the leaf nodes. This assumption means that each request will be responded and the user will not leave the edge leaf's coverage before the response arrives.    
\end{itemize}

In the absence of a caching scheme in a network, users submit their request for a specific type of data (we refer to as a file hereinafter) to the local edge node.
The submitted request will then traverse its way up to the parent node and then the primary source to evoke a response from the corresponding IoT device. In contrast, when a caching scheme is deployed, after receiving a request, the edge node will look up its cache memory for the requested file. If the file is available, the edge node will respond to the user itself and prevent further communications. Otherwise, the edge node will pass the request to the parent node, and now, the parent will go through the same process of checking its cache memory. In this way, the data source needs to generate and transmit a response only when there is no valid response available in the edge or the parent nodes. It is readily evident that 
the response time would be significantly shorter by avoiding multiple hops of communication. Besides, when data sources are less frequently forced to wake, generate and transmit a response, they will consume less energy.

Without loss of generality, we consider $N$ IoT devices that have the role of data sources and each of them can only produce one type of data. 
Given a file $i$, denoted as $f_i$, a ternary tuple $\{id_i, t_{gen}^i, t_{life}^{i} \}$ is introduced to represent the unique ID, the corresponding generation timestamp, and lifetime of that file. Lifetime is the duration of time that the content is still valid after being generated. Each type of data has its own lifetime, e.g., smoke detector’s data might be valid for just twenty seconds, because it is crucial to act upon a sudden change of smoke, while humidity data might be valid for hours. Freshness, i.e., $freshness_i$, is another metric that can measure the performance of a caching scheme and 
is defined as $freshness_i = ( t_{current} - t_{gen}^i) / t_{life}^{i} $, where $t_{current}$ is the current time. When a file reaches its expiration time, i.e., $freshness_i \geq 1$, the file is no longer valid and will be deleted from the cache memory.  

\subsection{Popularity Distribution}
\label{popdis}
Different from other works \cite{8713465, 9144224} that include multiple edge nodes (with one-layer architecture) for cooperative edge caching by exchanging cache information, we have considered multiple caching nodes to form a multi-layered networking architecture. The motivation behind the proposed hierarchical caching architecture is the regional-different popularity distribution\footnote{Regions are defined as the area that the edge nodes can cover and popularity refers to the frequency with which a file has been requested.}. The popularity of files follows different distributions from region to region. It is possible for a file $f_i^{\mathcal{X}}$ to be moderately popular in some regions $\mathcal{X}$ but not popular enough to be cache-worthy. None of the edge nodes caches this file in such a scenario and thus sends every request for $f_i$ to the parent node to be fetched from its source. However, from the parent node's point of view, things are different; the parent receives the accumulated requests for $f_i$ and might find out that $f_i$ is a cache-worthy file. In other words, the popularity distribution in the parent node is unique to itself, and thus we can also benefit from caching in the parent node.

Aside from being a more realistic system model, our empirical results (see Section \ref{simresult} for the detail) show that it is more beneficial to have a caching scheme in both edge and parent nodes rather than only at the edge nodes (even assuming an equal sum of memory space). 


\section{Problem Formulation}
\label{probfor}
In this section, we first explain how we formulate the caching problem to a Markov Decision Process (MDP) and then introduce the ingredients of solving the MDP problem, including the design of state, action, and reward function.

\subsection{Markov Decision Process Modeling}
Conventionally, at each time step $n$, MDPs are represented by an information tuple of $ \{s_n,a_n, p (s_{n+1}|s_n,a_n),r_n\} $ that respectively denote the current state $s_n \in \mathcal{S}$, chosen action $a_n \in \mathcal{A}$, a distribution function $p (s_{n+1}|s_n,a_n) \in \mathcal{P(S, A)}$ which gives the probability of a next state $s_{n+1}$ given the current state and action, and finally a reward $r_n$ that evaluates how good an action $a_n$ is in the state $s_n$. Implementing a DRL algorithm means that we are seeking a policy, in our case a caching policy $\pi(a_n|s_n)$, which gives the probability of choosing an action $a_n$ in the state $s_n$. It is intuitively clear that besides the immediate reward, caching decisions can also influence the rewards of the later steps, so it is only reasonable to consider long-term returns while making decisions. To do so, we define a cumulative reward $G_n$ as below
 
\begin{equation}
\label{Gn}
    G_n = \sum_{\tau=0}^T \gamma^t r_{n + \tau},
\end{equation}
where $n$ refers to time step, $r_\tau$ is the received reward in time step $\tau$, $T$ is the final time step. $\gamma \in (0, 1]$ is the discount factor that determines the effect of future rewards to current caching decisions. A smaller values of $\gamma$ emphasize on the short-term rewards and the greater values emphasize on the long-term returns.

The ultimate goal here is to find the optimal policy $\pi^*$, under which the expected cumulative reward $ \mathbb{E}[G_n | \pi]$ can be maximized
\begin{equation}
    \pi^* = \underset{\pi}{\mathrm{argmax}} \enspace \mathbb{E}[G_n | \pi].
\end{equation}

To find the optimal policy, we define the value function $V^\pi (s_n)$ and the action-value function $Q^\pi (s_n,a_n)$, respectively. $V^\pi (s_n)$ denotes the expected cumulative reward when starting from the state $s_n$  and following the policy $\pi$ after that. The action-value function $Q^\pi (s_n, a_n)$, also referred to as $Q$ function, denotes the expected cumulative reward starting from the state $s_n$, taking the action $a_n$, and after that following the policy $\pi$. Using Bellman equations \cite{textbookRL}, we can write these two functions as in (3) and (4), \begin{equation}
    V^\pi (s_n) = \sum_{a_n\in\mathcal{A}} \pi(a_n|s_n) \sum_{s_{n+1} \in \mathcal{S}} p(s_{n+1}|s_n,a_n) [r_n + \gamma V^\pi (s_{n+1})],
\end{equation}

\begin{align*} 
    Q^\pi (s_n,a_n) & = \sum_{s_{n+1} \in \mathcal{S}}  p(s_{n+1}|s_n,a_n)  \\
   & \cdot  \Bigg[r_n+ \gamma \sum_{a_{n+1}\in \mathcal{A}} \pi(a_{n+1}|s_{n+1})  Q^\pi (s_{n+1},a_{n+1}) \Bigg].
\end{align*} 

If any of these two functions is known, we can find the definite optimal policy $\pi^*$. The reason is that in each state $s_n$ we would know what action will yield the greatest immediate reward $r_n$ and the expected future value $V(s_{n+1})$; and we can simply define the optimal policy as choosing the most rewarding action in each time step. To obtain an optimal policy through value function or $Q$-function is known as value-based reinforcement learning approach.
However, traditional RL is limited to the conditions that state space is fully-observable and with low dimensions. However, considering the freshness feature of IoT data makes the IoT system more complex and dynamic. Therefore the prerequisites such as handcrafted features or low-dimensional search spaces (e.g., $\mathcal{S}$ and/or $\mathcal{A}$) is unavailable. Besides, the transition function $\mathcal{P(S,A)}$ is also unknown, making it difficult to obtain or even approximate those value functions. 

To conquer the above challenges, we leverage the deep reinforcement learning technique to approximate the value function or the Q function in Section \ref{ppo}. One potent tool that helps us approximate those functions is the neural networks (NNs). NNs let us avoid manual feature extraction and train the agents on raw (and potentially high-dimensional) observations instead. Moreover, by deploying neural  networks, it is possible to parametrize the policy $\pi_\theta (a_n|s_n)$ (a.k.a. $\pi_\theta$), with $\theta$ being the set of parameters. Then we can directly tune the parameters of the policy function in search of the optimal one based on the gradient of some performance measure with respect to the policy parameters $\theta$. 

\subsection{State, Action, and Reward}
Through interactions with the environment, the RL agent receives multiple observations (i.e., samples of the state space), and after analysing the observation and remembering from previous experiences, the agent then takes its action, which gradually achieves its ultimate goal, maximizing the value function. Since the freshness of transient data is a key factor in IoT caching problem, the corresponding consideration in terms of the state space and reward function are shown. In what follows, we describe the ingredients of MDP, including the space of State $\mathcal{S}$ and Action $\mathcal{A}$, as well as the design of reward function.

\subsubsection{State $\mathcal{S}$}
In order to form our observations, we first define the memory status of each caching node to be a part of state-space. Memory status is a three-row matrix representation for all cached files $\mathcal{M} = \{1,2, \cdots, M \}$ on this node. Each column 
corresponds to a file, e.g., $f_i$, and
contains the file ID, the freshness value as calculated in (\ref{fresh}), and the total cache hits $k_i$ for that file in the cache memory.


\begin{equation}
    MemStatus= \begin{bmatrix}
    id_1 & ... & id_M\\
    freshness_1 & ... & freshness_M\\
    k_1 & ... & k_M
    \end{bmatrix},
\end{equation}

\begin{equation}
\label{fresh}
freshness = \frac{ t_{current} - t_{gen}} {t_{life}}.
\end{equation}

If there is no content stored in a memory at some time slot, the file's ID, freshness and cache hit value are all set to zero. The ID of \text{requested} file (e.g., $\hat{f_i}$) and its lifetime account for the other part of the observation. We put these two values together as the request variable
\begin{equation}
    ReqVariable= \begin{bmatrix}
    id, & t_{life}
    \end{bmatrix}.
\end{equation}

In this way, the RL agent can observe whether the requested file is available in the cache memories or not. The freshness of stored items is also available, so the agent might decide to replace a file that is going to expire in the near future with the requested file.

\subsubsection{Action $\mathcal{A}$}
Assume a user has asked for a file $f$, and neither the corresponding edge node nor the parent has the file in its cache memory, and thus, the response should be fetched from the source node. In this case, the networking nodes, i.e., parent and edge nodes, should make a caching decision. There are two possible actions and the action space is $\mathcal{A} = \{0,1\}$. If $a=1$, the node will cache the file at hand; otherwise, the file will not be stored. If a caching node decides to store a file when there is enough free space to do so, the process is a straightforward saving action, but if the cache memory is full, the new file has to replace an existing file. In the latter case, the proposed caching strategy will point out which file has to be replaced \footnote{We clarify that the proposed caching scheme is not a proactive approach, i.e., no file will be cached prior to its corresponding request as in \cite{vanet3d, ma2020age}. In dynamic networks such as IoT, files are valid for a short period of times and their values can be constantly changing, proactive caching does not offer enough value and might add an extra burden to the backhaul links, which deviates from the purpose.}. 

\subsubsection{Reward}
\begin{figure}[t]
\centering
\resizebox{0.3\textwidth}{!}
{\includegraphics{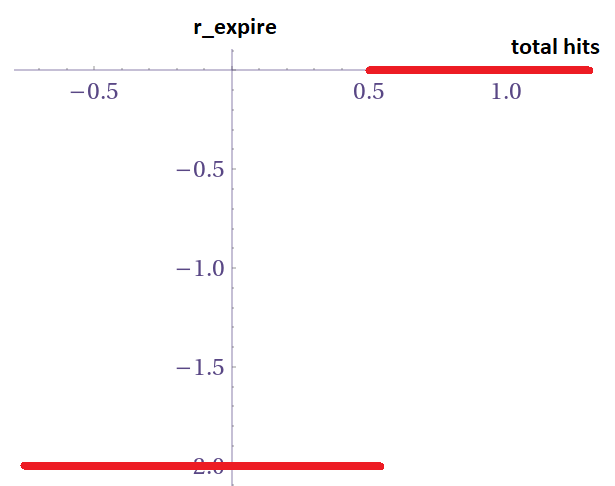}}
\caption{$r_{expire}$ function vs total number of hits ($C_1 = 1)$}
\label{Gr: r1}
\end{figure}

As we mentioned earlier, 
data may have limited lifetime in IoT networks, which raises concerns about data freshness\footnote{Expired data is worthless to respond to users' requests and should be evacuated from cache memory. It is readily evident that repeatedly caching a file with a limited lifetime cannot be very beneficial, even though it might be fairly popular. Moreover, storing files with a long lifetime and inferior popularity rank is not efficient either since it is very possible that the file will be expired without receiving any request during its lifetime.}.  As an example for the importance of freshness measure, a DRL agent might deduce that since a file is reaching its expiration, it can be a beneficial decision to replace it with the new and coming data. However, we should also note that freshness does not offer any intrinsic value; files are valid before expiration and a fresher version of the same file does not offer any advantage. To reflect the above points in our work, we \textit{keep the freshness of cached files in the observation, but we do not include it in the reward function}. However, for each file in the cache memory that expires without being used at least once, we give a negative reward (punishment) to the agent for storing that file. We keep track of the number of times a cached file has been hit; thus, we can calculate the reward function defined in (\ref{r-expire}) for each file in the cache memory after it expires. 

\begin{equation}
\label{r-expire}
    r^i_{expire} = [sign(k_i - 0.5)-1] \times C_1,
\end{equation}
where $sign (x)$ is the function that returns 1 if $x\geq 0$ and -1 otherwise. $C_1$ is the constant we choose as the punishment coefficient. Note that the total number of hits is an integer, thus by using (\ref{r-expire}) we are giving a negative reward only if a file has a total number of hits equal to zero. Refer to Figure \ref{Gr: r1} for a better illustration.

The ultimate objective of any good caching policy is to improve the cache hit rate, i.e. the rate at which users' requests are fulfilled locally by the caching nodes and consequently the communication cost is alleviated (less energy consumption). As such, increasing cache hit rate translates into shorter response time and higher energy efficiency. Delay (i.e. response time) is the time duration that users have to wait for their response. The longer distance a request should traverse to evoke a response, the longer the user should wait. Thus, it is most desirable if the first networking node has the requested file in its caching memory because the delay would be minimum. So, in order to encourage higher cache hit rates, we use the following reward function


\begin{equation}
    r_{hit} =  \frac{K \times C_2}{t_{current}},
\end{equation}

\begin{equation}
    K = \sum_{t=0}^{t_{current}} \sum_{i=1}^M k_i,
\end{equation}
where $ t_{current} $ is the current time step, $K$ is the sum of all the cache hits up until the current time step, and $C_2$ is a constant hyper-parameter which we choose arbitrarily. The final reward function is represented as follows
\begin{equation}
    r = r_{hit} + \sum_{i=1}^M r^i_{expire}.
\end{equation}

\section{Proximal Policy Optimization Solver}
\label{ppo}

In this section, we have deployed one version of the Actor-Critic (AC) models named Proximal Policy Optimization (PPO) algorithm \cite{ppo}. This algorithm is developed based on the Trust Region Policy Optimization (TRPO) algorithm \cite{trpo} and aims to inherit the data efficiency and reliability of the TRPO while avoiding its second-order optimizations. 

\begin{figure}[!t]
\centering
\resizebox{0.5\textwidth}{!}
{\includegraphics{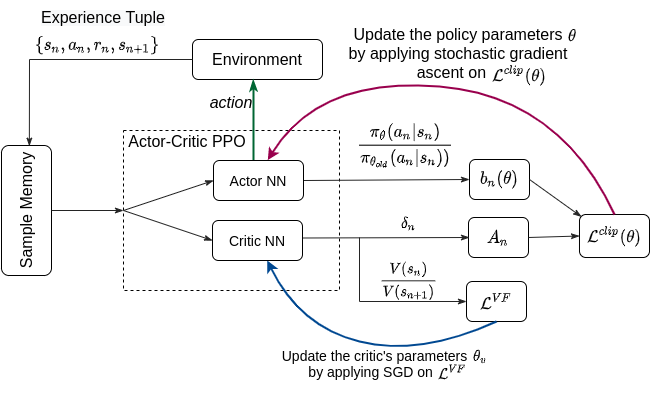}}
\caption{PPO with Actor-Critic style}
\label{Gr: PPO_actor_critic}
\end{figure}

The PPO algorithm in this work is based AC method, by which we can simultaneously approximate the policy and value function using two separate neural networks named actor and critic, denoted as $\theta$ and $\theta_v$, respectively. The actor decides which action should be taken and the critic evaluates the action produced by the actor by computing the value function $V^{\pi_\theta} (s_n; \theta_v )$. A brief description of how AC works is as follow.\\
\indent Step 1: Sample $\{s_n, a_n\}$ using policy $\pi_{\theta}$ from actor network; Execute $a_n$ and observe $s_{n+1}, r_n$; \\
\indent Step 2: Evaluate the advantage function (a.k.a.  TD error $\delta_n$ from critic network) by (\ref{advantage_function})
\begin{equation}
\label{advantage_function}
    A_n = r_n + \gamma V^{\pi_\theta}(s_{n+1}; \theta_v )- V^{\pi_\theta}(s_n; \theta_v ).
\end{equation}
\indent Step 3: Update the policy (i.e., $\theta$) via gradient ascent method, based on gradient of advantage function in Step 2.

Different from traditional AC, PPO optimizes a clipped Surrogate objective function to update the policy $\theta$, which is defined as

\begin{equation}
\label{clipped_sur}
    \mathcal{L}^{clip}(\theta) = \E_n[min\{b_n(\theta)A_n, clip(b_n(\theta),1-\epsilon,1+\epsilon)A_n\}], 
\end{equation}

\begin{equation}
\label{importance_sampling}
    \hspace{0.6cm} b_n(\theta) = \frac{\pi_\theta(a_n|s_n)}{\pi_{\theta_{old}}(a_n|s_n)},
\end{equation}
where $min$ gives the minimum argument given two inputs, $clip(\cdot)$ function takes in the probability ratio $b_n (\theta)$ defined in (\ref{importance_sampling}) but clips its value to be no more than $1+\epsilon$ and no less than $1-\epsilon$, with $\epsilon$ being a hyperparameter, $\theta_{old}$ in (\ref{importance_sampling}) is the vector of policy parameters before the update, and $A_n$ is the advantage function defined in (\ref{advantage_function}).

Finally, to update the parameters $\theta$, we use the following rule
\begin{equation}
\label{policy_update}
    \theta \leftarrow \underset{\theta}{\mathrm{argmax}} \enspace  \mathcal{L}^{clip}(\theta). 
\end{equation}
where $\mathrm{argmax}$ is usually achieved by the stochastic gradient ascent \cite{ppo}.

Algorithm \ref{ppo_alg} gives a more comprehensive elaboration on the details of the PPO algorithm. In order to update the critic network, we first need the cumulative rewards $G_n$ as defined in (\ref{Gn}). Then as in step 6, we apply stochastic gradient descent on the mean squared error and update the critic network's parameters $\theta_v$. Putting Algorithm \ref{ppo_alg} and Figure \ref{Gr: PPO_actor_critic} together should give us a clear image on how the PPO performs.




\begin{algorithm}[t!]
\caption{PPO Algorithm, Actor-Critic Style}
\begin{algorithmic}[1]
\label{ppo_alg}
\renewcommand{\algorithmicrequire}{\textbf{Input:}}
\renewcommand{\algorithmicensure}{\textbf{Output:}}
\REQUIRE initial NN parameters for actor and critic ($\theta$ and $\theta_v$)
\FOR {$iteration = 1,2, ...$}
\STATE Collect set of trajectories $\mathcal{T}=\{\tau_j\}$ by running the policy $\pi_\theta$ in the environment for $T$ times
\STATE Estimate advantage values $\hat{A}_1 ,..., \hat{A}_T$ 
\STATE Receive the rewards and compute $\hat{G}_1,..., \hat{G}_T$ 
\STATE Optimize surrogate loss w.r.t. $\theta$ (e.g., stochastic gradient ascent with Adam) \\
\hspace{2.3cm} $\theta \leftarrow \underset{\theta}{\mathrm{argmax}} \enspace  \mathcal{L}^{clip}(\theta)$
\STATE Fit value function by regression on mean squared error using stochastic gradient descent\\
\hspace{2.3cm} $\theta_v \leftarrow \underset{\theta_v}{\mathrm{argmin}} \enspace \mathcal{L}^{VF}$\\
where\\
$\hspace{1.25cm} \mathcal{L}^{VF} = \frac{1}{|\mathcal{T}|T} \sum_{\tau \in \mathcal{D}} \sum_{n=0}^T \big( V_{\theta_v}(s_n) - \hat{G}_n \big)^2 $
\ENDFOR
\end{algorithmic}
\end{algorithm}

\section{Simulations \& Results}
\label{result}
In order to conduct our experiment, we have used Python version 3.7, NumPy \cite{numpy}, TensorFlow 1.14.0 \cite{tf}, Keras, OpenAI's Gym \cite{gym}, and Stable Baselines \cite{SB} libraries\footnote{Stable Baselines is a library of reinforcement learning algorithm implementations based on OpenAI Baselines.}. We ran the simulation on the workstation with an Intel Core i7-9700 CPU and 16G RAM. Below, we first elaborate our simulation setup, including the popularity distribution of data, parameter setting of the proposed DRL solver, and the benchmarks for comparison (i.e., LRU, LFU, and DRL in \cite{mainref}) in Section \ref{simset}. The superiority of the proposed IoT caching design. in terms of the cache hit rate and energy consumption is provided and discussed in Section \ref{simresult}. In addition, we also demonstrate how our proposed design outperforms benchmarks when different average freshness and popularity of cached files are present.

\subsection{Simulation Setup}
\label{simset}
For ease of illustration, we consider a hierarchical IoT caching structure with one parent node, two edge nodes, and one hundred IoT devices producing the data.
Each IoT device only produces one type of file with a unique ID and a lifetime, which is sampled randomly from a uniform distribution with a minimum of 2 and a maximum of 14 time-steps.
The popularity distribution follows Zipf's law, which means that a file $f$ with a popularity-rank of $x$ is going to be requested with a probability of
\begin{equation}
\label{zipf}
    \bar p(x;\alpha,F) = \frac{x^{-\alpha} }{\sum_{x=1}^{F} x^{-\alpha}}, 
\end{equation}
where $F$ is the total number of files in the network, and $\alpha$ is the characterizing parameter to the distribution, which we call the skewness factor in this work. As we assign greater values to the $\alpha$, the request frequency gap between the most popular and least popular files will grow larger.
We have done our experiments with 24 different settings, resulting from four different request rate values $w$ and six different values of popularity skewness factor $\alpha$. The request rate of users $w$ varies from 0.5 to 2.0 requests per time step, and the popularity distributions have a varying $\alpha$ between 0.7 to 1.2. Please note that these distributions are unknown to the DRL agent; they are used only for generating requests in the simulations. We assume that each request can be fulfilled before its corresponding user leaves the network. There is no assumption for the users' arrival model since the DRL agent does not use any information except the ones received from the interactions with its environment to make caching decisions.     

The architecture of the DRL algorithm is depicted in Figure \ref{Gr: PPO_actor_critic}. The learning rate is set to 0.001, the discount factor $\gamma$ is equal to 0.99, and the agent will experience 16 steps before performing an update, and the number of training mini-batches per update is set to 4. The actor and critic NNs are feedforward neural networks, with two hidden layers and 64 neurons in each layer, and the activation function in use is $\tanh$. 

In this work, we compare our method with two of the best known conventional caching strategies: the least recently used (LRU) and least frequently used (LFU) algorithms. In addition, we have also implemented an existing DRL based caching method \cite{mainref} with a different reward function from ours to observe the effects of our proposed reward function.

Now, let us briefly review how each of these methods works:
\begin{enumerate}
    \item \textit{LRU}: In LRU, the stored files are always ranked based on how recently they have been used, and when a new file should replace one of the files in the memory, the least recently used file will be deleted from the memory. 
    \item \textit{LFU}: Similar to LRU, cached items are ranked, but the ranking criterion is the frequency of their requests. If the memory is full, the new file (will be cached) will replace the file with the least frequency of request.
    \item \textit{DRL}: In our work, we keep track of the number of hits of each file and use this information to design a reward function to encourage a high hit rate. On the other hand, related works such as \cite{mainref} directly consider the freshness of files in their reward function. To investigate the performance difference owing to the use of different reward functions, we have implemented the DRL approach in \cite{mainref} for comparison. We refer to this approach as  \textit{DRL} in our results.  
\end{enumerate}

In the following section, we have evaluated our proposed method based on the cache hit rate and energy consumption. We have also monitored the average freshness and popularity of cached files.

\subsection{Results and Discussions}
\label{simresult}
\begin{figure}[t]
\centering
\resizebox{0.45\textwidth}{!}
{\includegraphics{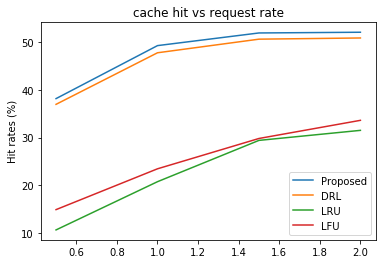}}
\caption{Hit rate vs varying request rates $w$}
\label{Gr: hit_w}
\end{figure}

\begin{figure}[t]
\centering
\resizebox{0.45\textwidth}{!}
{\includegraphics{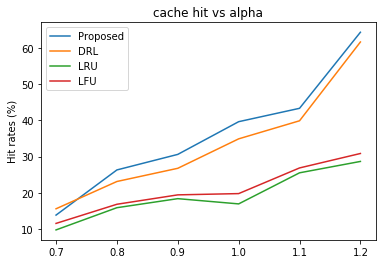}}
\caption{Hit rates vs popularity skewness $\alpha$}
\label{Gr: hit_a}
\end{figure}

Figures \ref{Gr: hit_w} and \ref{Gr: hit_a} show the cache hit rate versus varying request rate and popularity skewness factor, respectively. It is a clear trend that by increasing those two factors, the cache hit rate will increase as well. Higher request rate gives more chances to the cached files to be requested, and higher skewness factor makes the popularity gap between files even larger, which makes DRL agent  easier to learn the pattern. We can see that our proposed method tremendously outperforms the well-known LRU and LFU methods, and it does a better job than the existing DRL method with a different reward function. The proposed method starts with about 38\% cache hit, and it goes up to more than 52\%. The other DRL approach follows the proposed one with about 1.3\% gap, whereas the conventional methods LFU, which outperforms LRU, shows a cache hit range of 16\% to 34\% for different request rates. The cache hit rate versus the skewness factor has a similar outcome for different approaches.

\begin{figure}[t]
\centering
\resizebox{0.45\textwidth}{!}
{\includegraphics{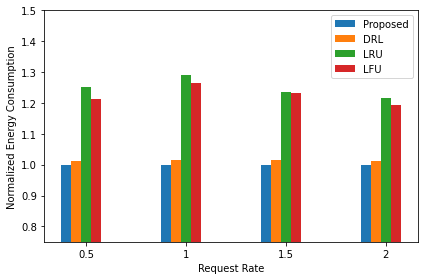}}
\caption{Energy consumption vs request rates $\omega$}
\label{Gr: energy_w}
\end{figure}

\begin{figure}[t]
\centering
\resizebox{0.45\textwidth}{!}
{\includegraphics{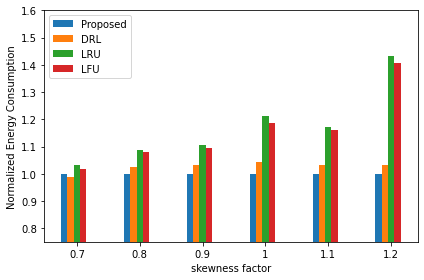}}
\caption{Energy consumption vs popularity skewness $\alpha$}
\label{Gr: energy_a}
\end{figure}

We have also simulated the energy consumption of the IoT devices for each caching scheme. Figure \ref{Gr: energy_w} and Figure \ref{Gr: energy_a} depict the simulation results with varying request rate and popularity skewness factor, respectively. We have normalized the energy consumption rates by our proposed method; values greater than 1 indicate that the energy consumption is higher than our proposed method for that specific configuration, and values less than one denote less energy consumption. We observe that our method uses the least energy for IoT devices among its counterparts. This result meets our expectation that the energy consumption is inversely proportional to the cache hit rate. The higher the cache hit rate, the lower the energy consumption.  

\begin{figure}[!h]
\centering
\resizebox{0.45\textwidth}{!}
{\includegraphics{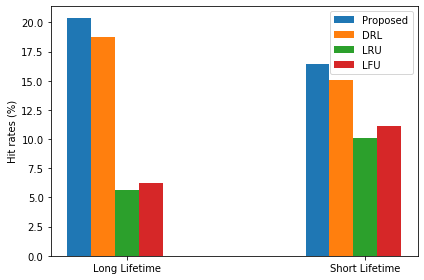}}
\caption{Hit rates vs lifetime }
\label{Gr: hit_pop}
\end{figure}

\begin{figure}[!h]
\centering
\resizebox{0.45\textwidth}{!}
{\includegraphics{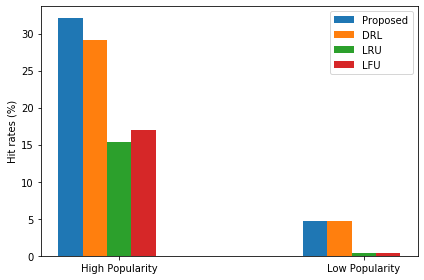}}
\caption{Hit rate vs popularity}
\label{Gr: hit_life}
\end{figure}

In order to better understand what makes the proposed algorithm to have a better cache hit rate, we took a look at the requests which have been responded to by a caching node. We categorize files based on their popularity and lifetime and then keep track of the requested files which have been hit in the cache memories. Figure \ref{Gr: hit_pop} and Figure \ref{Gr: hit_life} show these results which the average over all requests and over 24 different settings of our experiments (the 24 settings vary in request rate and the popularity skewness factor as shown in Table \ref{tab:freshness_table}). We can see that the DRL-based methods tend to store popular files, and also they favor files with a longer lifetimes, whereas conventional methods (i.e., LRU and LFU) might even favor files with shorter lifetimes. This shows that the DRL agent has learned that storing popular files with a longer lifetime is more beneficial, which is an entirely reasonable argument.

To evaluate the effects of the proposed reward function on the freshness of cached files, we have prepared Table \ref{tab:freshness_table}. This table shows the average freshness for all the cache memory files during the whole simulation for each of the 24 different settings. Note that a freshness value of 1 means the file has expired, and a freshness value of 0 means the file has just been generated. The lower the freshness value, the fresher the file is. In the last row of the table, we can observe the mean value of all those settings. Since in our proposed method, the reward function does not include any punishment for the higher value of freshness, the cached files have slightly older files in the memory compared to the other DRL method, which explicitly considers freshness in its reward function. Nevertheless, the gap between these two methods is marginal.

\begin{table}[t]
\begin{center}
\caption{Table of Average Freshness}
\begin{tabular}{|l|c|c|c|c|c|}
\hline
Setting & Proposed & DRL \cite{mainref}\\
\hline
$w=0.5, \alpha=0.7$  &    0.497 &   0.17\\
$w=0.5, \alpha=0.8$  &    0.494 &   0.329\\
$w=0.5, \alpha=0.9$  &    0.536 &   0.53\\
$w=0.5, \alpha=1.0$  &    0.496 &   0.495\\
$w=0.5, \alpha=1.1$  &    0.532 &   0.532\\
$w=0.5, \alpha=1.2$  &    0.527 &   0.523\\
\hline
$w=1, \alpha=0.7$  &    0.469 &   0.468\\
$w=1, \alpha=0.8$  &    0.457 &   0.457\\
$w=1, \alpha=0.9$  &    0.422 &   0.422\\
$w=1, \alpha=1.0$  &    0.451 &   0.450\\
$w=1, \alpha=1.1$  &    0.411 &   0.409\\
$w=1, \alpha=1.2$  &    0.433 &   0.433\\
\hline
$w=1.5, \alpha=0.7$  &    0.247 &   0.355\\
$w=1.5, \alpha=0.8$  &    0.355 &   0.355\\
$w=1.5, \alpha=0.9$  &    0.296 &   0.296\\
$w=1.5, \alpha=1.0$  &    0.346 &   0.346\\
$w=1.5, \alpha=1.1$  &    0.333 &   0.293\\
$w=1.5, \alpha=1.2$  &    0.324 &   0.307\\
\hline
$w=2, \alpha=0.7$  &    0.182 &   0.27\\
$w=2, \alpha=0.8$  &    0.279 &   0.279\\
$w=2, \alpha=0.9$  &    0.225 &   0.228\\
$w=2, \alpha=1.0$  &    0.271 &   0.275\\
$w=2, \alpha=1.1$  &    0.249 &   0.251\\
$w=2, \alpha=1.2$  &    0.293 &   0.237\\
\hline

\textbf{Average} & \textbf{0.380} & \textbf{0.362}\\
\hline
\end{tabular}
\end{center}
\label{tab:freshness_table}
\end{table}

\begin{figure}[!h]
\centering
\resizebox{0.45\textwidth}{!}
{\includegraphics{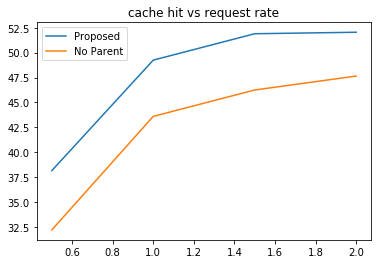}}
\caption{Comparison of hierarchical architecture (proposed) and flat architecture (edge caching only). The figure shows the hit rate vs request rates $\omega$}
\label{Gr: NP_w}
\end{figure}

\begin{figure}[!h]
\centering
\resizebox{0.45\textwidth}{!}
{\includegraphics{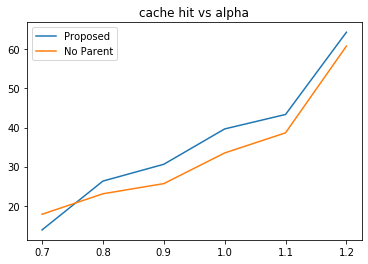}}
\caption{Comparison of hierarchical architecture (proposed) and flat architecture (edge caching only). The figure shows the hit rate vs the popularity skewness}
\label{Gr: NP_a}
\end{figure}

Finally, in order to see the benefits of a hierarchical model architecture, we have implemented a single-layer caching scheme where there are two edge nodes capable of caching, and there is no parent node in the network. For the sake of fair comparison, we keep the sum of memory capacity of these two edge nodes on par with the cumulative memory capacity of the three nodes (one parent node and two edge nodes) in our hierarchical architecture. Figure \ref{Gr: NP_w} and Figure \ref{Gr: NP_a} show that the proposed hierarchical architecture can lead to higher cache hit rate and consequently lower energy consumption. The reason is that the parent node in the hierarchical architecture can effectively cache the files, which are commonly prevalent in multiple regions
but might not be cache-worthy by edge nodes, thus increasing the overall cache hit rate.

\section{Conclusion}
\label{conclu}
In this work, we have considered an IoT network with transient data files. Taking into account the limited lifetime of data and the energy consumption of IoT devices, we have formulated the caching problem into an MDP problem and then tackled the MDP problem with a deep reinforcement learning method using a proximal policy optimization algorithm. We have proposed a hierarchical architecture for implementing the DRL caching algorithms in two layers. The proposed method does not require any prior knowledge about the environment, users, or requests. The extensive experimental results show that our proposed DRL method and hierarchical architecture can make smarter decisions in caching, significantly improve the cache hit rate, and reduce the energy consumption of the IoT devices in comparison to benchmark caching schemes including LRU and LFU, and an existing DRL solution.

\bibliographystyle{IEEEtran}
\bibliography{reference}

\begin{thebibliography}{10}
\providecommand{\url}[1]{#1}
\csname url@samestyle\endcsname
\providecommand{\newblock}{\relax}
\providecommand{\bibinfo}[2]{#2}
\providecommand{\BIBentrySTDinterwordspacing}{\spaceskip=0pt\relax}
\providecommand{\BIBentryALTinterwordstretchfactor}{4}
\providecommand{\BIBentryALTinterwordspacing}{\spaceskip=\fontdimen2\font plus
\BIBentryALTinterwordstretchfactor\fontdimen3\font minus
  \fontdimen4\font\relax}
\providecommand{\BIBforeignlanguage}[2]{{%
\expandafter\ifx\csname l@#1\endcsname\relax
\typeout{** WARNING: IEEEtran.bst: No hyphenation pattern has been}%
\typeout{** loaded for the language `#1'. Using the pattern for}%
\typeout{** the default language instead.}%
\else
\language=\csname l@#1\endcsname
\fi
#2}}
\providecommand{\BIBdecl}{\relax}
\BIBdecl

\bibitem{myConf}
A.~Nasehzadeh and P.~Wang, ``A deep reinforcement learning-based caching
  strategy for internet of things,'' in \emph{Proc. IEEE/CIC International
  Conference on Communications in China (ICCC)}, 2020, pp. 969--974.

\bibitem{iotStuff}
A.~Nordrum, ``Popular internet of things forecast of 50 billion devices by 2020
  is outdated,'' \emph{IEEE Spectrum}, Aug 2016.

\bibitem{6225433}
U.~Niesen, D.~Shah, and G.~W. Wornell, ``Caching in wireless networks,''
  \emph{IEEE Transactions on Information Theory}, vol.~58, no.~10, pp.
  6524--6540, 2012.

\bibitem{lfu}
D.~Lee, J.~Choi, J.-H. Kim, S.~H. Noh, S.~L. Min, Y.~Cho, and C.~S. Kim,
  ``Lrfu: A spectrum of policies that subsumes the least recently used and
  least frequently used policies,'' \emph{IEEE transactions on Computers},
  vol.~50, no.~12, pp. 1352--1361, 2001.

\bibitem{ccVelipasalar}
C.~Zhong, M.~C. Gursoy, and S.~Velipasalar, ``A deep reinforcement
  learning-based framework for content caching,'' in \emph{Proc. 52nd Annual
  Conference on Information Sciences and Systems (CISS)}, 2018, pp. 1--6.

\bibitem{8308415}
M.~Meddeb, A.~Dhraief, A.~Belghith, T.~Monteil, and K.~Drira, ``How to cache in
  icn-based iot environments?'' in \emph{Proc. IEEE/ACS 14th International
  Conference on Computer Systems and Applications (AICCSA)}, 2017, pp.
  1117--1124.

\bibitem{7883826}
S.~Wang, X.~Zhang, Y.~Zhang, L.~Wang, J.~Yang, and W.~Wang, ``A survey on
  mobile edge networks: Convergence of computing, caching and communications,''
  \emph{IEEE Access}, vol.~5, pp. 6757--6779, 2017.

\bibitem{textbookRL}
R.~S. Sutton and A.~G. Barto, \emph{Reinforcement learning: An
  introduction}.\hskip 1em plus 0.5em minus 0.4em\relax MIT press, 2018.

\bibitem{Qlearning}
V.~Mnih, K.~Kavukcuoglu, D.~Silver, A.~A. Rusu, J.~Veness, M.~G. Bellemare,
  A.~Graves, M.~Riedmiller, A.~K. Fidjeland, G.~Ostrovski \emph{et~al.},
  ``Human-level control through deep reinforcement learning,'' \emph{nature},
  vol. 518, no. 7540, pp. 529--533, 2015.

\bibitem{zhu2018deep}
H.~Zhu, Y.~Cao, W.~Wang, T.~Jiang, and S.~Jin, ``Deep reinforcement learning
  for mobile edge caching: Review, new features, and open issues,'' \emph{IEEE
  Network}, vol.~32, no.~6, pp. 50--57, 2018.

\bibitem{sadeghi}
A.~Sadeghi, G.~Wang, and G.~B. Giannakis, ``Deep reinforcement learning for
  adaptive caching in hierarchical content delivery networks,'' \emph{IEEE
  Transactions on Cognitive Communications and Networking}, vol.~5, no.~4, pp.
  1024--1033, 2019.

\bibitem{vanet3d}
Z.~Zhang, Y.~Yang, M.~Hua, C.~Li, Y.~Huang, and L.~Yang, ``Proactive caching
  for vehicular multi-view 3d video streaming via deep reinforcement
  learning,'' \emph{IEEE Transactions on Wireless Communications}, vol.~18,
  no.~5, pp. 2693--2706, 2019.

\bibitem{pre7Iintegratedbigdata}
Y.~He, N.~Zhao, and H.~Yin, ``Integrated networking, caching, and computing for
  connected vehicles: A deep reinforcement learning approach,'' \emph{IEEE
  Transactions on Vehicular Technology}, vol.~67, no.~1, pp. 44--55, 2017.

\bibitem{9310745}
S.~Nath and J.~Wu, ``Deep reinforcement learning for dynamic computation
  offloading and resource allocation in cache-assisted mobile edge computing
  systems,'' \emph{Intelligent and Converged Networks}, vol.~1, no.~2, pp.
  181--198, 2020.

\bibitem{pre8Jointfog}
Y.~Wei, F.~R. Yu, M.~Song, and Z.~Han, ``Joint optimization of caching,
  computing, and radio resources for fog-enabled iot using natural
  actor--critic deep reinforcement learning,'' \emph{IEEE Internet of Things
  Journal}, vol.~6, no.~2, pp. 2061--2073, 2018.

\bibitem{GreenDRL}
X.~He, K.~Wang, H.~Huang, T.~Miyazaki, Y.~Wang, and S.~Guo, ``Green resource
  allocation based on deep reinforcement learning in content-centric iot,''
  \emph{IEEE Transactions on Emerging Topics in Computing}, vol.~8, no.~3, pp.
  781--796, 2018.

\bibitem{QoEDRL}
X.~He, K.~Wang, and W.~Xu, ``Qoe-driven content-centric caching with deep
  reinforcement learning in edge-enabled iot,'' \emph{IEEE Computational
  Intelligence Magazine}, vol.~14, no.~4, pp. 12--20, 2019.

\bibitem{ma2020age}
M.~Ma and V.~W. Wong, ``Age of information driven cache content update
  scheduling for dynamic contents in heterogeneous networks,'' \emph{IEEE
  Transactions on Wireless Communications}, vol.~19, no.~12, pp. 8427--8441,
  2020.

\bibitem{8713465}
W.~Jiang, G.~Feng, S.~Qin, and Y.~Liu, ``Multi-agent reinforcement learning
  based cooperative content caching for mobile edge networks,'' \emph{IEEE
  Access}, vol.~7, pp. 61\,856--61\,867, 2019.

\bibitem{9144224}
Y.~Zhang, B.~Feng, W.~Quan, A.~Tian, K.~Sood, Y.~Lin, and H.~Zhang,
  ``Cooperative edge caching: A multi-agent deep learning based approach,''
  \emph{IEEE Access}, vol.~8, pp. 133\,212--133\,224, 2020.

\bibitem{federatedDRL}
X.~Wang, C.~Wang, X.~Li, V.~C. Leung, and T.~Taleb, ``Federated deep
  reinforcement learning for internet of things with decentralized cooperative
  edge caching,'' \emph{IEEE Internet of Things Journal}, vol.~7, no.~10, pp.
  9441--9455, 2020.

\bibitem{mainref}
H.~Zhu, Y.~Cao, X.~Wei, W.~Wang, T.~Jiang, and S.~Jin, ``Caching transient data
  for internet of things: A deep reinforcement learning approach,'' \emph{IEEE
  Internet of Things Journal}, vol.~6, no.~2, pp. 2074--2083, 2018.

\bibitem{MultiAgentCc}
C.~Zhong, M.~C. Gursoy, and S.~Velipasalar, ``Deep multi-agent reinforcement
  learning based cooperative edge caching in wireless networks,'' in
  \emph{Proc. IEEE International Conference on Communications (ICC)}, 2019, pp.
  1--6.

\bibitem{rana2021systematic}
B.~Rana, Y.~Singh, and P.~K. Singh, ``A systematic survey on internet of
  things: Energy efficiency and interoperability perspective,''
  \emph{Transactions on Emerging Telecommunications Technologies}, vol.~32,
  no.~8, p. e4166, 2021.

\bibitem{ppo}
J.~Schulman, F.~Wolski, P.~Dhariwal, A.~Radford, and O.~Klimov, ``Proximal
  policy optimization algorithms,'' \emph{arXiv preprint arXiv:1707.06347},
  2017.

\bibitem{trpo}
J.~Schulman, S.~Levine, P.~Abbeel, M.~Jordan, and P.~Moritz, ``Trust region
  policy optimization,'' in \emph{Proc. International conference on machine
  learning (ICML)}, 2015, pp. 1889--1897.

\bibitem{numpy}
C.~R. Harris, K.~J. Millman, S.~J. Van Der~Walt, R.~Gommers, P.~Virtanen,
  D.~Cournapeau, E.~Wieser, J.~Taylor, S.~Berg, N.~J. Smith \emph{et~al.},
  ``Array programming with numpy,'' \emph{Nature}, vol. 585, no. 7825, pp.
  357--362, 2020.

\bibitem{tf}
M.~Abadi, A.~Agarwal, P.~Barham, E.~Brevdo, Z.~Chen, C.~Citro, G.~S. Corrado,
  A.~Davis, J.~Dean, M.~Devin \emph{et~al.}, ``Tensorflow: Large-scale machine
  learning on heterogeneous distributed systems,'' \emph{arXiv preprint
  arXiv:1603.04467}, 2016.

\bibitem{gym}
G.~Brockman, V.~Cheung, L.~Pettersson, J.~Schneider, J.~Schulman, J.~Tang, and
  W.~Zaremba, ``Openai gym,'' \emph{arXiv preprint arXiv:1606.01540}, 2016.

\bibitem{SB}
A.~Raffin, A.~Hill, A.~Gleave, A.~Kanervisto, M.~Ernestus, and N.~Dormann,
  ``Stable-baselines3: Reliable reinforcement learning implementations,''
  \emph{Journal of Machine Learning Research}, 2021.

\end{thebibliography}

\end{document}